\documentclass{optica-article}

\journal{opticajournal} 

\articletype{Research Article}

\usepackage{lineno}
\usepackage{graphicx}
\usepackage{svg}
\nolinenumbers 

\begin{document}

\title{A simple, robust and cost-effective method to achieve dispersion matching in swept source OCT}

\author{Shau Poh Chong,\authormark{1,*} Peter T\"{o}r\"{o}k,\authormark{1,2,3,4}}

\address{\authormark{1}Singapore Centre for Environmental Life Sciences Engineering
(SCELSE), Nanyang Technological University, 60 Nanyang Drive,
637551, Singapore.\\
\authormark{2}School of Physical and Mathematical Sciences, Nanyang Technological
University, 50 Nanyang Drive, 639798, Singapore.\\
\authormark{3}Lee Kong Chian School of Medicine, Nanyang Technological University,
59 Nanyang Drive, 639798, Singapore.\\
\authormark{4}Institute for Digital Molecular Analytics and Science (IDMxS),
Nanyang Technological University, 59 Nanyang Drive, 639798, Singapore.}

\email{\authormark{*}josiah.chong@ntu.edu.sg} 


\begin{abstract*} 
Optical path length and dispersion matching in both measurement and reference arms of an OCT system is critical for  achieving  bandwidth-limited axial resolution. To minimize or eliminate dispersion mismatch, most, if not all,  fiber-based OCT realisations employ a reference arm configuration that is as closely identical to the measurement arm as possible. This typically includes a collimator, dispersion compensating material (or sometimes a set of lenses), as well as a mirror (or retro-reflector) mounted on a translation stage. However, this solution makes the total instrument cost higher and the setup bulkier than necessary and it also renders the reference arm mechanically unstable. Here, a simple yet robust, low-cost reference arm setup is presented and its ability to compensate for measurement arm dispersion is demonstrated. We use a single-mode fiber cleaved and polished perpendicular to the fiber axis to construct the reference arm. The length and material of the fibre is determined by  considering the optical path length and dispersion of the measurement arm. Experimental images demonstrate the operation of the novel reference arm in our Swept-source Optical Coherence Tomography.

\end{abstract*}

\section{Introduction}
In optical coherence tomography (OCT), dispersion matching of both arms of the interferometer is critical to ensure that the point spread function is optimal,  otherwise both axial resolution and signal-to-noise ratio (SNR) \cite{attendu2019} degrade. Dispersion mismatch happens as each constituent wavelength of the  light source experiences different optical path length as it traverses the optical system, including the sample in the measurement arm, as well as in the reference arm. When using broad optical bandwidth illumination to improve axial resolution of the OCT, dispersion mismatch is  exacerbated. To minimise or eliminate  dispersion mismatch within the optical interferometer, typically the sample and reference arms are constructed identically, which means that the an identical copy of all the optical components in the measurement arm is placed in the reference arm\cite{Murthy2025}. However, this adds significant cost and complexity to the  setup. Alternatively, a simpler hardware compensation schemes can be adopted by using glass plate \cite{Hitzenberger99} (such as BK7 or fused silica) to provide a fixed amount of dispersion in the reference arm; or by a prism pair \cite{Bouma1995, Drexler1999} that provides adjustable amount of dispersion. These solutions make the OCT reference arm unnecessary bulky, mechanically sensitive  and complicated to maintain. Another solution is to exploit changes in dispersion of stretched optical fibers \cite{Iyer2009}, however this approach depends heavily on the type and length of the fiber chosen and can only introduce small variations in dispersion.  

In principle, the dispersion mismatch can also be reduced by adjusting fiber length of the reference arm and using the light reflected back into the fibre from the glass/air interface of the core at the cleaved interface (typically around 4$\%$) as reference\cite{Juskaitis}. The use of reflected light from  the fibre end as reference has been demonstrated previously in common-path OCT\cite{Tan2009,liu2013}, however, for Mach–Zehnder Interferometer interferometer (split path) based OCT design this solution has not been implemented to the best of our knowledge. Such a strategy can greatly simplify the optical design of the reference arm as shown in this work. 

In the current work, we propose to employ a single mode fibre (SMF) with cleaved and polished fiber tip to generate the reference signal for  the interferometer. The length of the open-ended SMF was carefully path length matched to that of the measurement arm including the optical path length introduced by the objective lens. As anticipated such arrangement significantly simplifies the reference arm design and reduce the overall physical dimensions of the OCT setup. In addition our solution improves optical/thermal stability and it requires minimal maintenance/alignment. We have implemented the aforementioned method in our home-built 1.7 $\mu$m swept source OCT (SSOCT), which has the benefit of achieving higher penetration depth into highly-scattering samples such as biological tissues \cite{Chong2015}, mainly due to reduced multiple scattering from the sample at longer wavelengths.   

\section{Methods}\label{sec11}

We  constructed a conventional reference arm for our SSOCT,  consisting of a reflective collimator (RC02APC-P01, Thorlabs), a dispersion compensation block (LSM03DC, Thorlabs) to match the dispersion induced by the objective lens (LSM03, Thorlabs) placed in the measurement arm, as well as a mirror mounted on a manual translation stage to allow adjustment of the reference delay.  On the other hand, we prepared a FC/APC connectorised SMF which at one end was cleaved and polished to achieve a flat tip with the length of the fibre calculated such as to provide dispersion similar to that in measurement arm. We used these two compensation methods alternately to provide dispersion matching (Fig. \ref{fig:fig01}). 

\subsection{System setup}
\begin{figure}[h]
\centering
            \vspace{0.5mm}
            \includegraphics[width=0.905\linewidth]{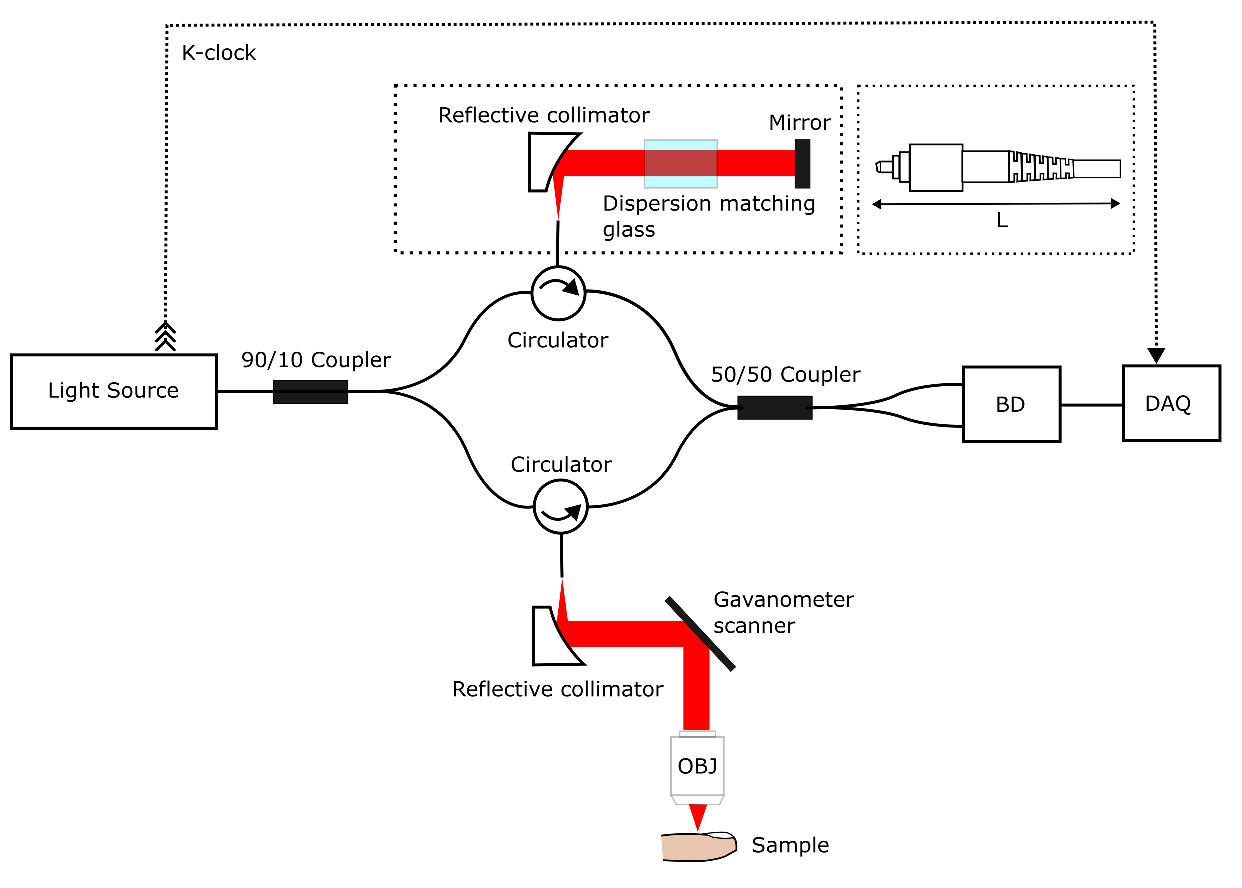}\hspace{1mm}
            \vspace{0.5mm}
            
\caption{Schematic of our swept source OCT system.  M: mirror, BD: balanced photodetector, and RC: reflective collimator. In our setup, both methods conventional reference arms setup, and cleaved fiber tip, are interchangeable for OCT imaging. }
\label{fig:fig01}
\end{figure}

Our OCT system design (see Fig. \ref{fig:fig01}) incorporates a commercial swept source (Santec, Inc., HSL-40-90-B, ) with central wavelength of 1700 nm and sweeping range of 135 nm.  The maximal output power of the laser is up to 35 mW with coherence length more than 10 mm.  The output light from the source was fed into a 90:10 coupler (TW1650R2A1, 1650 ± 100 nm, Thorlabs Inc.) and split into the measurement and reference arms by a pair of circulators (6015-3-APC, 1525 – 1601 nm, Thorlabs Inc.).  In both the measurement and reference arms the beam from the circulators was collimated by using a reflective collimator.  In the measurement arm the collimated beam was directed to a galvanometer scanner pair (Saturn 5B, Scannermax) to perform raster scanning.  An objective lens (LSM03, Thorlabs) with effective focal length of 36 mm was then employed to focus the beam onto the sample. The objective lens was placed at a near telecentric configuration such that the chief ray exiting the objective lens was approximately parallel to the optic axis.  The use of reflective optics in the setup ensured that these elements did not introduce longitudinal chromatic aberration which are inherent in most refractive optics based broad optical bandwidth OCT.  A high-speed balanced photodetector (Santec, Inc., BPD-200-HS-1.7) with operation wavelength range of 1200 – 2300 nm was employed to detect the interference signal which was digitized by a high-speed analogue/digital converter (ADC - Santec, Inc., HAD-5200B-S). A custom LabVIEW program provided by Santec Inc was modified for triggering, synchronization, data acquisition and real time processing.  Specifically, the DAQ board (HAD-5200B-S, Santec, Inc) comes with an FPGA for real time SS-OCT image processing including real-time resampling and dispersion compensation. The swept source laser is also equipped with an integrated auxiliary interferometer to generate K-clock signal for wavenumber linear sampling. Together with the A-line trigger (termed as A-clock here), generated by the swept-source, both the A-clock and K-clock were fed into the DAQ for synchronization and sampling of the OCT signals. The spectral interferogram was detected by the balanced detector and sampled at 4096 points per A-scan using the digitiser within the DAQ board. The A-clock was also used to synchronise the XY galvanometer scanner, such that each period of each A-trigger clock corresponds to pixel dwell time of the scanner. The analogue output voltages for the XY galvanometer scanner were then output from the DAQ board to the scanner driver. In our setup, the wavenumber-sampled interferograms without dispersion compensation were saved as raw data.  Spectral Gaussian shaping were then performed to reduce spurious peaks in the OCT signals and numerical dispersion compensation was performed on all A-line data before Fast Fourier Transformation (FFT) of the interferogram to obtain the cross-sectional OCT images. The conventional reference arm contains a reflective collimator, a dispersion compensating block (Thorlabs LSM03DC, thickness of 17.8 mm; N-SK4 glass and refractive index of 1.59) and a manual translation stage mounted reference mirror. The dispersion compensating block is sold for the LSM03 objective lens and is designed to compensate for the dispersion of the LSM03 lens.

\begin{figure}[h]
\centering

            \vspace{0.5mm}
            \includegraphics[width=0.705\linewidth]{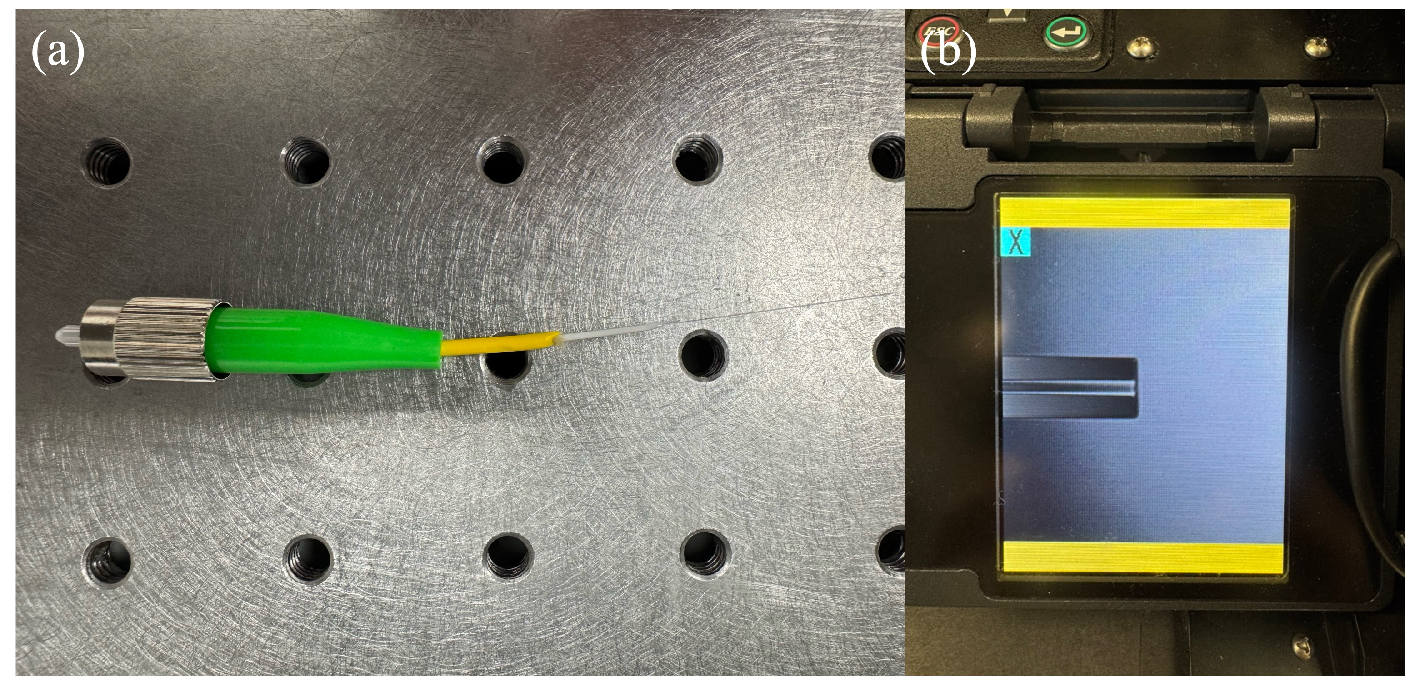}\hspace{1mm}
            \vspace{1.0mm}

\caption{(a) A photo showing the cleaved SMF with the length around 111 mm. (b) A photo showing the cleaved end of the fiber is flat to ensure optimal reflectivity from the air-glass interface on the fiber tip.}
\label{fig:fig06a}
\end{figure}

As shown in Fig. \ref{fig:fig01}, our novel fiber-based reference arm employs a low cost commercial FC/APC connectorised single-mode-fiber (SMF, OS2 9/125 $\mu$m, FS Singapore) developed for telecommunication application. The length of the SMF was determined by first measuring the physical length of the measurement arm of our SSOCT including the distance from the reflective collimator to the focal plane of the objective lens and then calculating the optical path length from refractive index data provided by the manufacturers' datasheet.   The total length of the measurement arm is 163 mm and so based on the refractive index of the SMF at 1.4684, the length of the cleaved fiber should be 111.21 mm. With this information, we cut the SMF to around this length, removed the outer jacket using a blade, and subsequently cleaved it using an optical fiber cleaver (CT50, Fujikura). The cleaved SMF is shown in the photo in Fig. \ref{fig:fig06a}a. We also verified that the cleaved end is clean and flat using a commercial splicer (as shown in Fig. \ref{fig:fig06a}b). 

\subsection{System characterization}
\begin{figure}[h]
\centering
            \vspace{0.5mm}
            \includegraphics[width=0.555\linewidth]{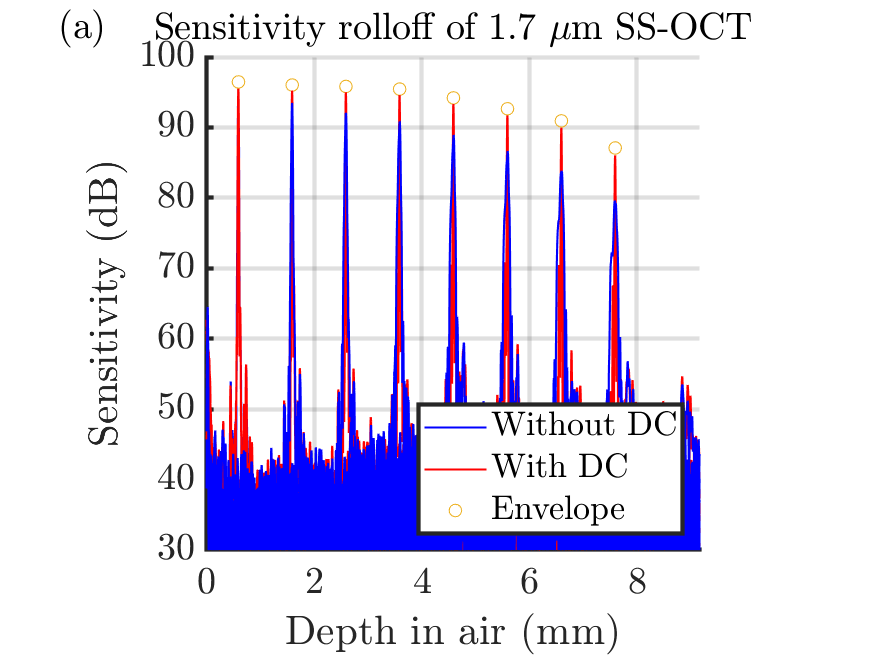}\hspace{-6mm}
            \includegraphics[width=0.4175\linewidth]{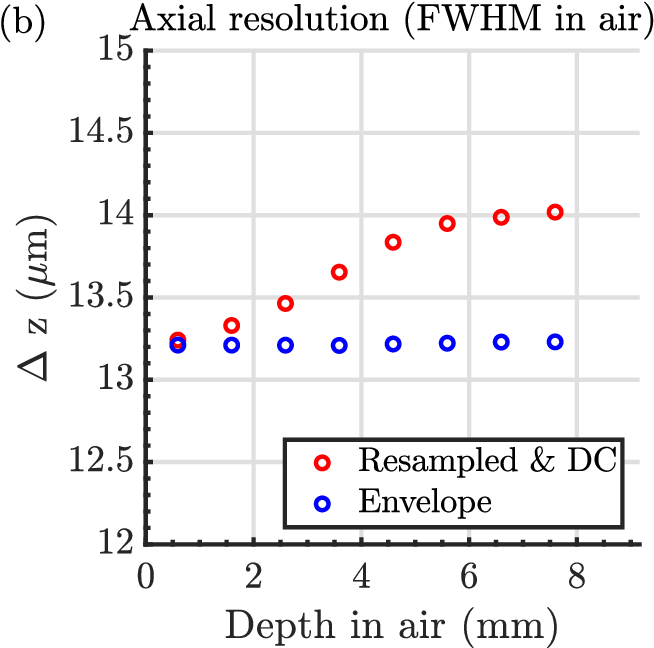}\hspace{1mm}
            \vspace{1.0mm}

\caption{(a) Sensitivity rolloff of the 1.7 $\mu$m SS-OCT system measured using reflection from a mirror, both with and without dispersion compensation (DC) is shown. (b) Measured axial resolution of the reflection point spread function from a mirror.  Similar to the rolloff, the measured axial resolution approaches that set by the coherence envelope, indicating correct k-resampling and dispersion compensation.}
\label{fig:fig02}
\end{figure}

Using the conventional reference arm setup, the imaging range of the system was measured to be approximately 9.56 mm. To measure the roll-off characteristics of the laser, we employed the conventional reference arm setup (which constitutes a reflective mirror on a translation stage) that allowed us to vary the reference arm length. By measuring the interference fringe at different mirror positions, and subsequent A-line processing, one can obtain the axial point-spread-function (PSF) of the mirror, the roll-off curve can then be obtained (Fig. \ref{fig:fig02}a). The axial resolution can then be measured by estimating the 3 dB width of the measured PSF in log scale (i.e. full-width-half-maximum in linear scale) which is shown in Fig. \ref{fig:fig02}b. The sensitivity roll-off of the system was measured to be $\approx$ 5dB for the first 5.6 mm of the total imaging range (see Fig. \ref{fig:fig02}(a)), providing an achievable axial resolution of less than 14 $\mu$m (in air) for a similar imaging range (see Fig. \ref{fig:fig02}(b) ).  The actual resolution when imaging turbid media and for biological samples was scaled by the refractive index, around 1.35. 

\begin{figure}[!h]
\centering
            \vspace{0.5mm}
            \includegraphics[width=0.4275\linewidth]{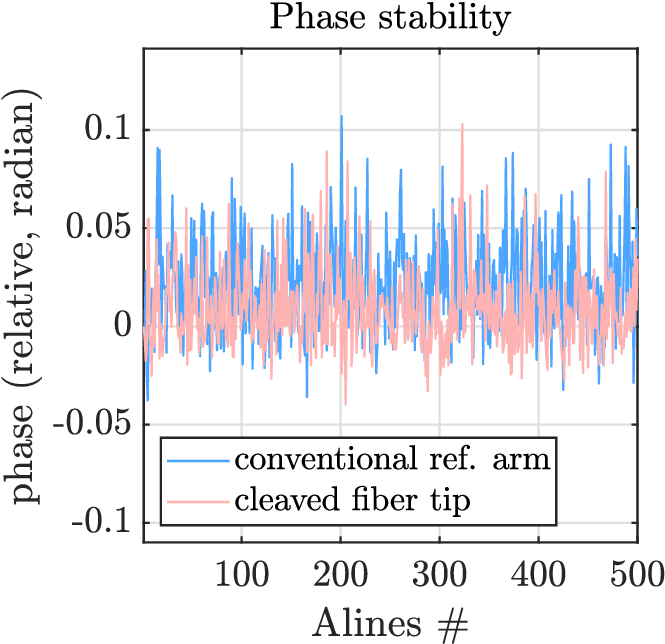}\hspace{1mm}
            \vspace{0.5mm}
            \vspace{0.5mm}
            
\caption{Phase stability measured by analyzing the phase fluctuation of the OCT interference signals from both methods conventional reference arms setup, and cleaved fiber tip, compared in the current study.}
\label{fig:fig03}
\end{figure}

To compare the phase stability of the conventional and proposed methods, a silver mirror was placed at the focus of the objective lens to generate an OCT interference signal. The length of the reference arm in the conventional setup was adjusted so that the interference signals were of similar frequency. In both cases, 10,000 A-lines were acquired sequentially, in M scan. We then extracted the phase information from the interference signals from both cases and plotted the maximum deviation of their corresponding phases. As shown in Fig. \ref{fig:fig03}, both methods achieved similar phase stability, with standard deviation of the phase fluctuations measured to be 0.0240 and 0.0205 radians, for conventional reference arm setup and cleaved fiber tip, respectively. 

\section{Results and discussion}

\begin{figure}[!h]
\centering
            \vspace{0.5mm}
            
            \includegraphics[width=0.965\linewidth]{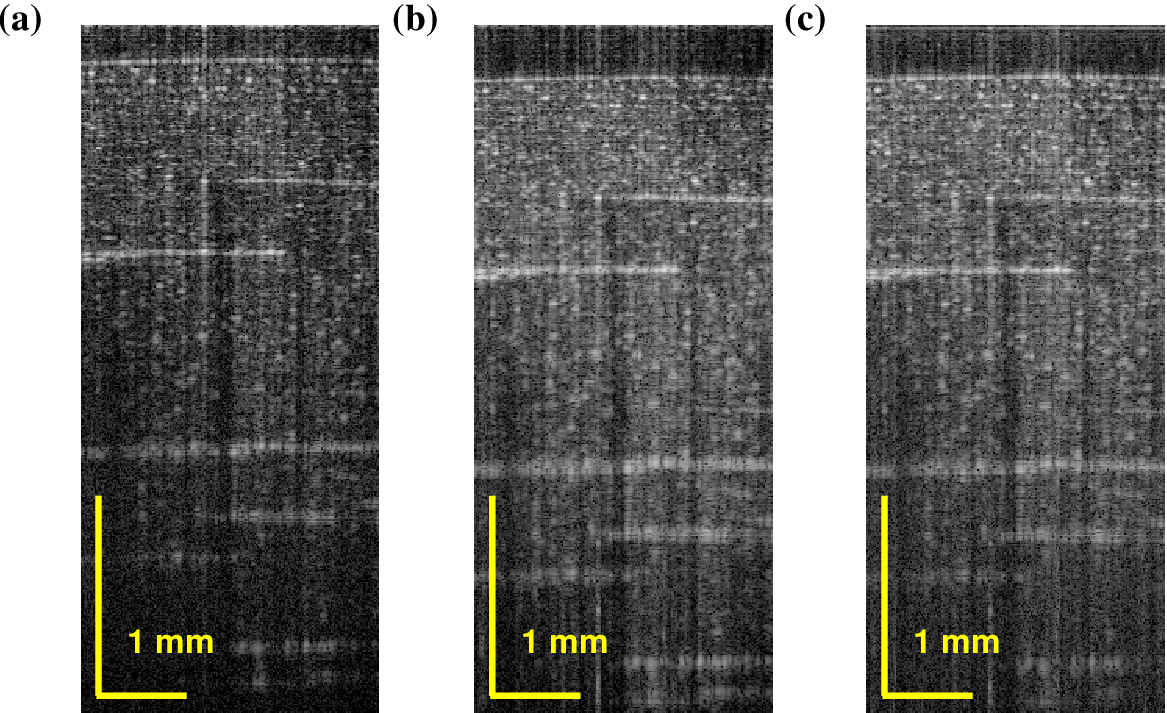}\hspace{1mm}
            
            \vspace{0.5mm}
            \vspace{0.5mm}
            
\caption{Comparison of SSOCT images of layers of Scotch\textsuperscript{TM} transparent tapes using (a) conventional reference arm and (b) cleaved SMF, both without spectral reshaping and further numerical dispersion compensation. OCT image of panel (b) but after Gaussian spectral shaping (c) and numerical dispersion compensation.}
\label{fig:fig0a}
\end{figure}

We then employed our SSOCT to image layers of Scotch\textsuperscript{TM} transparent tape as a simple validation experiment. In Fig. \ref{fig:fig0a}, the images (using the same grayscale level) captured using (a) conventional reference arm and (b) cleaved SMF show almost identical image contrast, albeit slightly higher SNR for FR. The higher SNR in (b) is simply due to the higher amount of light reflected back to the interferometer. It is worth mentioning that the conventional reference arm constructed using the retro-reflector, dispersion compensation block, and mirror requires frequent alignment to reflect enough reference power in contrast to the cleaved SMF which is alignment-free. Both images in Fig. \ref{fig:fig0a} (a) and (b) are processed without spectral reshaping and numerical dispersion compensation. The image in panel (c) shows the same data as (b) after spectral reshaping. It is evident in (a) - (c) that the depth-dependent dispersion degrades the image contrast at the deeper section of the image. We then applied the numerical dispersion compensation approach described in \cite{Wojtkowski2004}, and the features in the deeper regions of the image (\ref{fig:fig01}d) appear sharper at the cost of the shallower regions becoming blurred. This is because the numerical dispersion described in \cite{Wojtkowski2004} only performed the average dispersion and not that of the entire depth in the sample. Numerical depth-dependent dispersion compensation \cite{Mikkel2018,Dai2017, Kho2019} has previously been demonstrated, but over a smaller depth range. As the imaging depth of SSOCT usually covers more than a few mm as demonstrated here, a more robust depth-dependent dispersion approach covering the entire depth would be required. This will be further investigated in future work. Though the application of the proposed bare cleaved fibre is favourable in terms of cost, size and stability as compared to a conventional OCT system reference arm, a bare cleaved surface could be prone to contamination and damage, thus requiring careful handling. Hence, a protective shield (similar to endoscopic catheter) or encapsulation might be required to minimize the need for regular maintenance and improve long-term robustness. 

\begin{figure}[!h]
\centering
            \vspace{0.5mm}
            
            \includegraphics[width=0.665\linewidth]{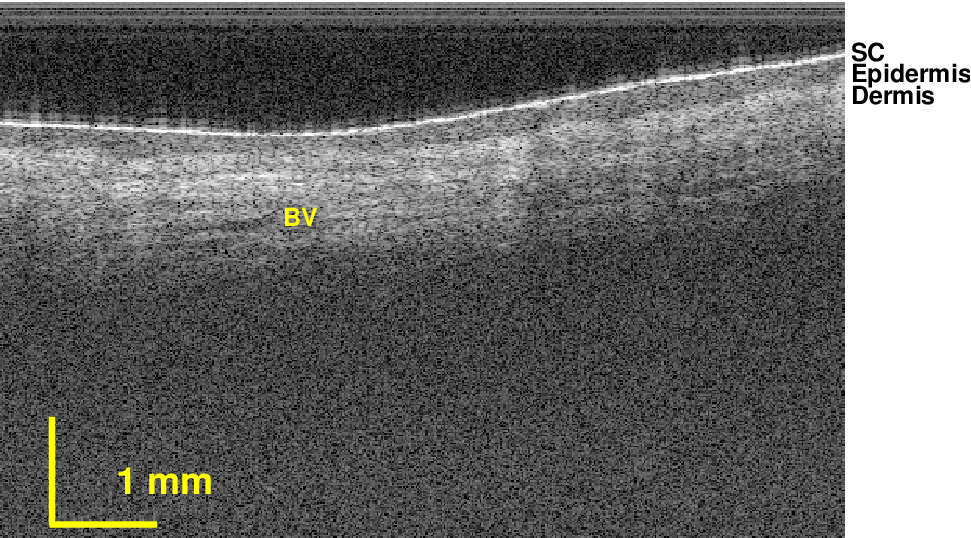}\hspace{1mm}
            
            \vspace{0.5mm}
            \vspace{0.5mm}
            
\caption{SSOCT image of human skin, specifically the back of the finger. A very bright layer corresponds to stratenum corneum, followed by dimmer layer of epidermis and then dermis layer are visible. Blood vessel (BV) could also be observed within the dermis layer, characterized by signal void due to high light attenuation of blood. The penetration depth into the normal skin is more than 1 mm.  SC: stratum corneum layer; BV: blood vessel. }
\label{fig:fig5}
\end{figure}

The imaging performance of the SSOCT with the new dispersion matching method was also validated by imaging healthy human skin \cite{Wan2021, Chong2020}. Before imaging, 75$\%$ alcohol wipe was used to disinfect and clean the skin surface to be imaged. In this study healthy hands skin was imaged from 2 subjects between 35 and 45 years of age. The imaging was carried out in a closed climate controlled room, at 22$^{\circ}$C and with 55$\%$ relative humidity. The optical power used for human skin imaging was set below 3 mW. A representative B scan image of the back of the finger is shown in Fig. \ref{fig:fig5}. Penetration depths of up to 1.5 mm were achieved and good contrast of the layered structures of the human skin was observed. Notably, a very bright stratum corneum layer, a dimmer layer of epidermis and a brighter layer of dermis is visualized. A blood vessel within the dermis layer could also be observed, characterized by signal void due to high light attenuation of blood. However due to the relatively poor axial resolution of the current SSOCT, i.e. $\approx$ 10 $\mu$m within tissue, finer structures are not readily visible \cite{Oliver2018}.

\section{Conclusion}\label{sec13}

We have demonstrated a novel simple, robust, and low-cost approach to perform matching the measurement arm dispersion in the reference arm in a  swept source OCT. The SMF employed is commercially available, and the cost of the is very low, i.e. a few US dollars, and is significantly lower than the components (inclusive of collimator, dispersion compensation block, mirror, attenuation block and translation stage etc) that construct the conventional reference arm; thus presenting an attrative alternative for the reference arm module for SSOCT. The new proposed method leverages on the relatively large axial imaging range, which is about 10 mm (in air) for system, as well as the negligible sensitivity rolloff of the SSOCT, and is more suitable for application where adjustment of the reference pathlength is not necessary.  The performance of the method proposed in this work was compared to that of an OCT using conventional reference arm setup by imaging layers of Scotch\textsuperscript{TM} transparent tape, as well as human skin. We showed that, in terms of performance, our approach is comparable to traditional reference arm design. Our method has the advantage of simpler and more compact swept source OCT design, with the added benefit of being alignment-free, mechanically stable and having low maintenance. Future plan includes modulating the reflectivity of the cleaved fiber tip for optical sensitivity as well as integrating it to OCT endoscopy for cost efficiency and optimal dispersion compensation. 

\begin{backmatter}
\bmsection{Funding}
We appreciate funding from Singapore A*AME YIRG grant A1884c0018 and National Research Foundation Singapore (NRF-CRP24-2020-0077).

\bmsection{Acknowledgment}
The authors appreciate Dr. Daiqi Xiong and Prof. Wonkeun Chang from the NTU School of Electrical $\&$ Electronic Engineering for their advice and expertise in the matter related to fiber cleaving. We also appreciate Masahito Hoshikawa from Santec Japan for assistance with the 1.7 $\mu$m swept source.

\bmsection{Disclosures}
The authors declare no conflict of interest.

\bmsection{Data Availability Statement} Data presented in this paper are not publicly available at this time but may be obtained from the authors upon reasonable request.

\end{backmatter}


\bibliography{Optica-template}






\end{document}